# Constraints on the Alfvénicity of Switchbacks

O.V. Agapitov[1,2], J. F. Drake[3,4], M. Swisdak[3], K.-E. Choi[2], N. Raouafi[5]

**ABSTRACT**

Switchbacks (SBs) are localized structures in the solar wind containing deflections of the magnetic field direction relative to the background solar wind magnetic field. The amplitudes of the magnetic field deflection angles ($\theta$) for different SBs vary from ~40 to ~160-170 degrees. Alignment of the perturbations of the magnetic field ($\Delta \vec{B}$) and the bulk solar wind velocity ($\Delta \vec{V}$) is observed inside SBs, so that $\Delta \vec{V} \sim \Delta \vec{B}$ when the background magnetic field is directed toward the sun (if the background solar wind magnetic field direction is anti-sunward then $\Delta \vec{V} \sim -\Delta \vec{B}$, supporting anti-sunward propagation in the background solar wind frame). This causes spiky enhancements of the radial bulk velocity inside SBs. We have investigated the deviations of SB perturbations from Alfvénicity by evaluating the distribution of the parameter $\alpha$, defined as the ratio of the parallel to $\Delta \vec{B}$ component of $\Delta \vec{V}$ to $\Delta \vec{V}_A = \Delta \vec{B}/4\pi n_i m_i$ inside SBs, i.e. $\alpha = V_{\parallel}/|\Delta \vec{V}_A|$ ($\alpha = |\Delta \vec{V}|/|\Delta \vec{V}_A|$ when $\Delta \vec{V} \sim \Delta \vec{B}$), which quantifies the deviation of the perturbation from an Alfvénic one. Based on Parker Solar Probe (PSP) observations, we show that $\alpha$ inside SBs has systematically lower values than it has in the pristine solar wind: $\alpha$ inside SBs observed during PSP Encounter 1 were distributed in a range from ~0.2 to ~0.9. The upper limit on $\alpha$ is constrained by the requirement that the jump in velocity across the switchback boundary be less than the local Alfvén speed. This prevents the onset of shear flow instabilities. The consequence of this limitation is that the perturbation of the proton bulk velocity in SBs with $\theta > \pi/3$ cannot reach $\alpha = 1$ (the Alfvénicity condition) and the highest possible $\alpha$ for a SB with $\theta = \pi$ is 0.5. These results have consequences for the interpretation of switchbacks as large amplitude Alfvén waves.

## 1. INTRODUCTION

A recent major discovery of Parker Solar Probe (PSP, Fox et al. 2016) was the presence of large numbers of localized velocity spikes associated with magnetic structures containing sudden deflections in the local radial magnetic field at 35.7-50 solar radii (RS) near the first PSP perihelion (Bale et al. 2019; Kasper et al. 2019, Dudok de Wit et al. 2020; Larosa et al. 2021, Raouafi et al. 2023). The observed rotation angle inside these structures varies from a few degrees up to a full reversal of the radial magnetic field component, which inspired their designation as "switchbacks" (SB). The time duration of a SB from the PSP data varies over a wide range from tens of seconds to tens of minutes (Dudok de Wit et al. 2020), which suggests that their size does not depend on the characteristic plasma scales. The alignment of the perturbations of the magnetic field ($\Delta \vec{B}$) and bulk solar wind velocity ($\Delta \vec{V}$) inside SBs (so that $\Delta \vec{V} \sim \Delta \vec{B}$ for the background magnetic field directed toward the sun) causes spiky enhancements of the radial bulk velocity inside SBs (Bale et al. 2019; Kasper et al. 2019). These enhancements attracted attention to these structures as a potential source of solar wind acceleration and heating (Bale et al. 2023). If the background solar wind magnetic field direction is anti-sunward then $\Delta\vec{V} \sim -\Delta\vec{B}$, supporting anti-sunward propagation in the background solar wind frame. Such alignment suggested a possible connection with Alfvénic waves as the possible source of SB generation and formation (Squire et al. 2020, 2022). The possible relation of SBs to magnetic flux ropes was proposed by Drake et al. (2021) and Agapitov et al. (2022) to explain the often-observed temperature enhancement, deviation from Alfvénicity, and composition structure of magnetic field perturbations inside SBs.

[1] Corresponding author agapitov@ssl.berkeley.edu
[2] Space Sciences Laboratory, University of California, Berkeley, CA 94720
[3] University of Maryland, College Park, MD, USA
[4] Institute for Research in Electronics and Applied Physics, University of Maryland, College Park, MD
[5] Applied Physics Laboratory, Johns Hopkins University, Laurel, MD, USA

We present results of an investigation of the geometry, de Hoffman-Teller frame parameters, and deviation from Alfvénicity of SBs based on PSP observations collected during the first solar perihelion (Encounter 1) and the SB database, which contains a list of 306 SBs. During Encounter 1, PSP was nearly co-rotating with the Sun for more than one week and was immersed in a slow but highly Alfvénic solar wind emerging from a small equatorial coronal hole (Kasper et al. 2019; Bale et al. 2019; Badman et al. 2020), making the data particularly useful for a statistical study of SB properties. Because the SBs observed during Encounter 1 were likely connected to similar source conditions, the distance from the Sun (i.e. the evolution time) and the background solar wind parameters were the factors that determined the differences in the observed SB properties (Mozer et al. 2020)

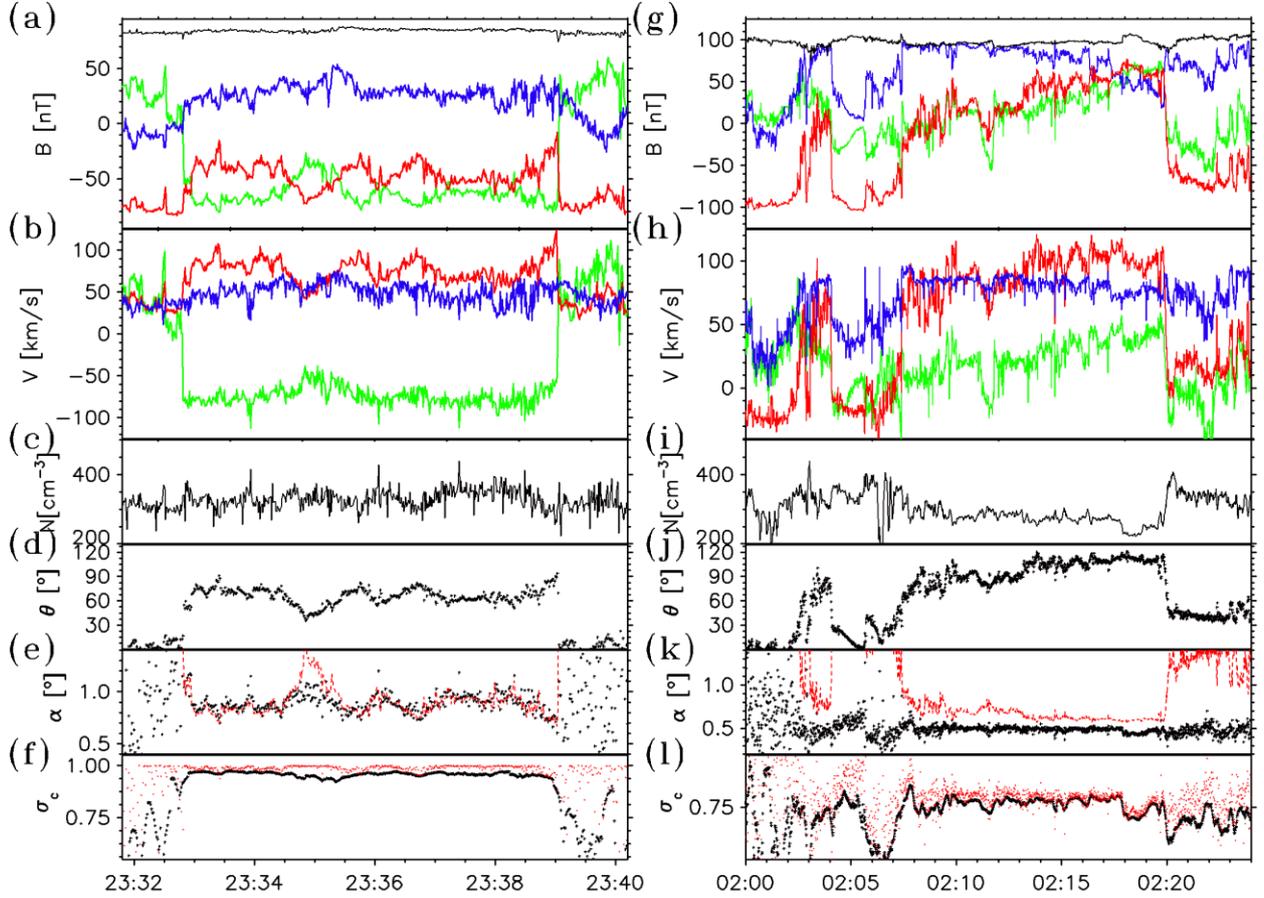

**Figure 1.** Two switchbacks encountered by PSP on November 6, 2018: (a) – magnetic field components in the RTN coordinate system (R-red; T-blue; N-green). The magnitude of the magnetic field is shown by the black curve; (b) – proton bulk velocity components in the RTN coordinate system (the same color coding as for panel (a)); (c) – proton density; (d) – the angle of magnetic field rotation $\theta$ (relative to the magnetic field $\vec{B}_{SW}$ recorded before the SB); (e) – the Alfvénicity parameter $\alpha$ ($|\Delta\vec{V}|/|\Delta\vec{B}|(4\pi n_i m_i)^{-1/2}$) (for aligned $\Delta\vec{V}$ and $\Delta\vec{B}$). The red points indicate the geometrical limit for $\alpha$: $\alpha_c = |\vec{B}_{SW}|/|\vec{B}_{SB} - \vec{B}_{SW}|$ based on the assumption that $|\Delta\vec{V}|$ is limited

by the local value of $V_A$. (f) – the cross-helicity $\sigma_c$. The red points indicate perfect alignment of the magnetic field and proton bulk velocity perturbations, so that $\sigma_c$ depends only on $\alpha$ as $\sigma_{c\alpha} = 2\alpha/(1 + \alpha^2)$. Panels (g-l) present the second switchback parameters in the same format.

## 2. DATA DESCRIPTION, PROCESSING TECHNIQUE, AND THE RESULTS

We use here data from PSP Encounter 1 (Bale et al. 2016; Kasper et al. 2016; Case et al. 2016) along with a database of 309 SBs determined by visual inspection to be intervals of deflected (relative to the background solar wind $\vec{B}_{SW}$) magnetic field $\vec{B}_{SB}$ defined by sharp boundaries (sharp in comparison with the structure duration). No threshold for angle of rotation $\theta$ was applied. Two magnetic field SBs recorded at about 36 Solar Radii (RS) from the Sun (November 6, 2018 during Encounter 1) are shown in Figure 1 and exhibit ~75-80 and ~100 degrees rotation of the magnetic field vector inside the structures (the magnetic field from FIELDS suite (Bale et al. 2016) and velocity components from Solar Wind Electrons Alphas and Protons (SWEAP, Kasper et al. 2016) are shown in the RTN coordinate system with R the radial direction directed from the Sun center, N the normal to the ecliptic plane component, and T the azimuthal component).

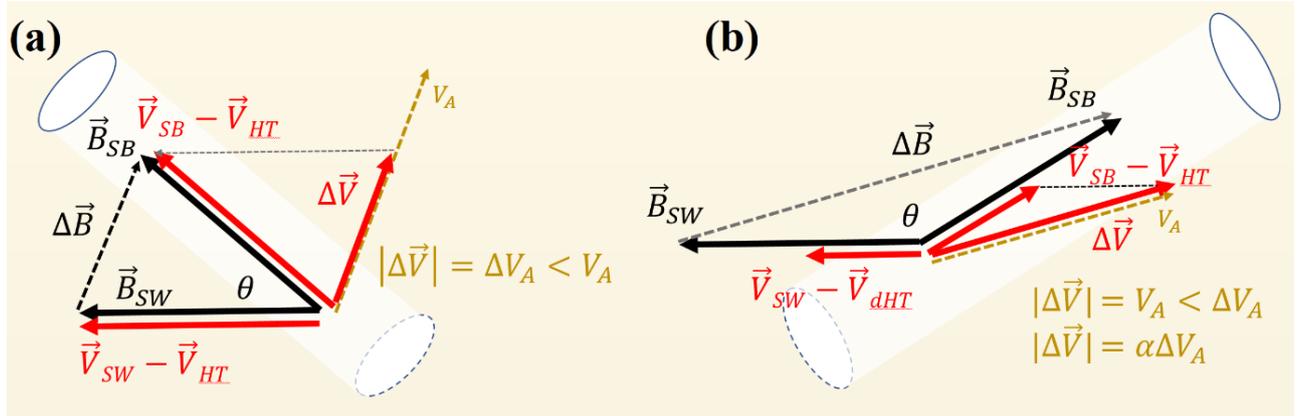

**Figure 2.** Schematic of the magnetic field (black) and plasma bulk velocity perturbation (red) in SBs in the de Hoffman-Teller frame (a) - for the Alfvenic perturbation, which is characterized by $\Delta\vec{V} = \Delta\vec{V}_A = \Delta\vec{B}/\sqrt{4\pi n}$, $|\vec{V}_{SB} - \vec{V}_{HT}| = V_A$, and $|\vec{V}_{SW} - \vec{V}_{HT}| = V_A$; and (b) – when the value $\Delta\vec{V}$ is limited by the local $V_A$. The velocity values are correspondingly scaled by the parameter $\alpha$: $\Delta\vec{V} = \alpha\Delta\vec{V}_A$, $|\vec{V}_{SB} - \vec{V}_{HT}| = \alpha V_A$, and $|\vec{V}_{SW} - \vec{V}_{HT}| = \alpha V_A$, with $\alpha = |\vec{B}_{SW}|/|\vec{B}_{SB} - \vec{B}_{SW}|$.

The sharp rotation of the direction of the magnetic field at the boundary while $\vec{B}$ remains nearly constant in magnitude ($|\vec{B}_{SW}| \approx |\vec{B}_{SB}|$), and the radial magnetic field change (up to changing sign as in Figure 1a-c), are typical characteristics of these events. The boundaries range in widths from tens of km (several proton inertial lengths) to thousands of km (Krasnoselskikh et al. 2020; Larosa et al. 2021; Mozer et al. 2020) with the scale range of SBs 10^4-10^5 km (Larosa et al. 2022). The perturbations of the proton bulk velocity $\vec{V}$ (Figure 1b and 1) follow the magnetic field $\vec{B}$ dynamics (Figures 1a and 1g) illustrating a high level of correlation between perturbations of the magnetic field and the proton bulk speed inside SBs ($\vec{B}_{SB}$ and $\vec{V}_{SB}$ respectively). The alignment of the $\vec{B}$ and $\vec{V}$ perturbations, $\Delta\vec{B} = \vec{B}_{SB} - \vec{B}_{SW}$ and $\Delta\vec{V} = \vec{V}_{SB} - \vec{V}_{SW}$, so that $\Delta\vec{V} \sim \Delta\vec{B}$ inside SBs, are seen in Figures 1a,b (SB1) and Figures 1g,h (SB2). This is shown schematically in Figure 2, which represents the geometry of SB1 and SB2 in Figure 2a and 2b respectively.

Plasma densities (from Solar Probe Cup measurements, Case et al. 2016) inside SBs are comparable with the background solar wind values, with fractional deviations only around $\pm 0.1$. Localized density enhancements at the switchback boundaries are often observed (Agapitov et al. 2020; and discussed in detail by Farrell et al. (2020)) and are presumably related to the ballistic propagation of SBs relative to the background solar wind speed. The geometrical characteristics of SBs are presented in Figure 1d-f (the variables are presented in the schematic in Figure 2). The location of the sharp change of the magnetic field angle $\theta$ defines the switchback boundaries. The average value is $70 \pm 5°$ for SB1 and $100 \pm 10°$ for SB2. It is convenient to quantify the Alfvénicity of a perturbation by introducing the parameter $\alpha = V_{||}/|\Delta\vec{V}_A|$ ($\alpha = |\Delta\vec{V}|/|\Delta\vec{V}_A|$ when $\Delta\vec{V} \sim \Delta\vec{B}$), where $V_{||}$ is the component of $\Delta\vec{V}$ along $\Delta\vec{B}$, and $\Delta\vec{V}_A = \Delta\vec{B}/\sqrt{4\pi n_i m_i}$. Figure 1e and 1h show some decrease of $\alpha$ inside SB1 and SB2. The critical value of $\alpha$ calculated based on the constraint that $\Delta\vec{V}$ be smaller than the local Alfven speed $V_A$ (so that the critical Alfvenic value is $\alpha_c = |\vec{B}_{SW}|/|\Delta\vec{B}|$) is shown by the red dots. For SB1, the values of $\alpha$ are between 0.8 and 1.1 and closely follow the variation of $\alpha_c$. For SB2, the values of $\alpha$ fall between 0.4 and 0.5, which is below $\alpha_c$, so the correlation between $\alpha$ and $\alpha_c$ as seen in SB1 was not observed. The variation of the normalized cross-helicity ($\sigma_c = 2\Delta\vec{V}_A \cdot \Delta\vec{V}/(|\Delta\vec{V}_A|^2 + |\Delta\vec{V}|^2)$) is presented in Figure 1f and Figure 1l. A general increase of $\sigma_c$ is observed inside SB1 and SB2. The red dots indicate the $\sigma_{c\alpha}$ values calculated from the values of $\alpha$ with the assumption that $\Delta\vec{B}$ and $\Delta\vec{V}$ are perfectly aligned so that $\sigma_{c\alpha} = 2\alpha/(1 + \alpha^2)$. The observed very good correspondence of $\sigma_c$ values to $\sigma_{c\alpha}$ inside switchbacks confirms the generally better alignment of $\Delta\vec{B}$ and $\Delta\vec{V}$ inside SBs than in the background solar wind. This supports applicability of the parameter $\alpha$ to quantify SB Alfvénicity.

The orientation of the magnetic field and plasma bulk velocity in the de Hoffman-Teller (dHT) frame is shown schematically in Figure 2. The dHT frame was calculated for the switchback-solar wind system. We used a transverse to the axis of the SB component of the dHT velocity ($V_{HT\perp}$) to estimate the velocity of switchback propagation in the solar wind frame. The SB axis is estimated as the average direction of magnetic field inside the SB (Krasnoselskikh et al. 2020). This transverse-to-the-axis-of-a-switchback component of the dHT velocity is compared with the transverse component of the plasma flow velocity inside the SB in Figure 3a, indicating the good agreement ($|\vec{V}_{SB} - \vec{V}_{SW}|_\perp = |\vec{V}_{HT} - \vec{V}_{SW}|_\perp$). The good correlation between the perpendicular velocity of SBs in the solar wind frame compared with the dHT velocity of the SW in Figure 3a establishes that there is a common dHT frame that encompasses both the SW and SBs. So, SBs can be considered in a form of the magnetic structure with the axial plasma flow (the plasma flow component along the averaged magnetic field inside the SB) and with transverse "drift" velocity in the solar wind frame.

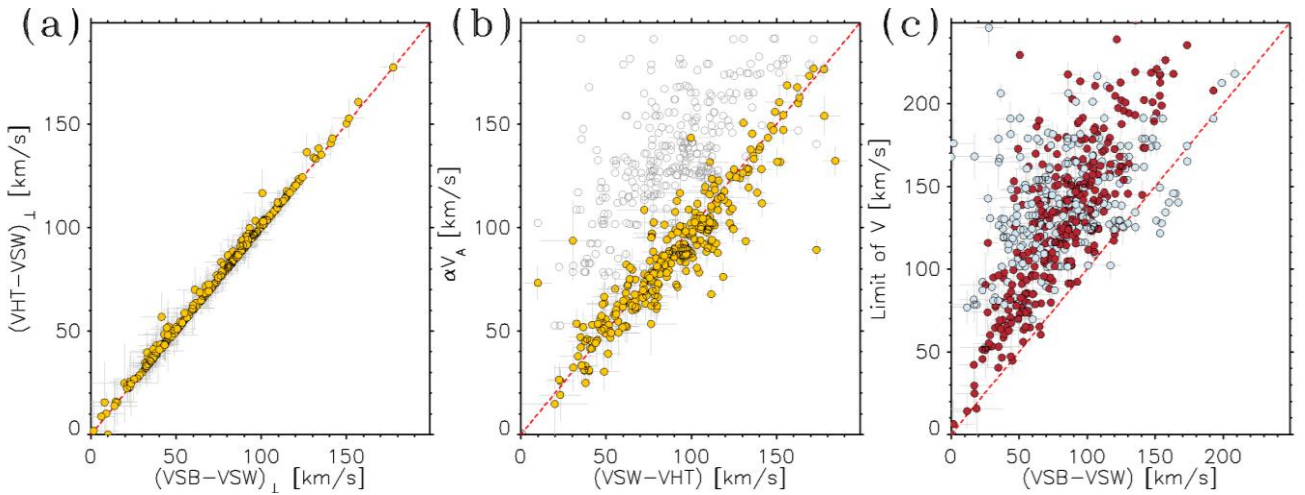

**Figure 3.** (a) - transverse to the SB axis component of dHT speed in the solar wind frame ($\vec{V}_{HT} - \vec{V}_{SW}$) versus the transverse component of the corresponding SB velocity ($\vec{V}_{SB} - \vec{V}_{SW}$). The SB axis is estimated as the average direction of magnetic field inside the SB (Krasnoselskikh et al. 2020). (b) – plasma flow velocity in the dHT frame $|\vec{V}_{SW} - \vec{V}_{HT}|$ versus $V_A$ and $\alpha V_A$ – better alignment with $\alpha V_A$ (the filled circles) than with $V_A$ (the empty circles) is observed. (c) - $V_A$ (the blue filled circles) and $\Delta V_A$ (the red filled circle) plotted versus $|\Delta V|$. Note that $|\Delta V| < \Delta V_A$ if $\Delta V_A < V_A$ so $\alpha$ can reach 1, and $|\Delta V| < V_A$ so $|\Delta V|$ is limited by $V_A$, so $\alpha < 1$ for this regime.

Because the magnetic field magnitude inside and outside the SB are similar, $|\vec{V}_{SW} - \vec{V}_{HT}| = |\vec{V}_{SB} - \vec{V}_{HT}|$ (see the schematic in Figure 2) and is equal to the local $V_A$ for the Alfvénic pertrubation. However, the constraint that $\Delta \vec{V}$ be smaller than the local $V_A$ leads to upper limits on the Alfvénicity parameter $\alpha$, as indicated by the dark yellow arrow in Figure 2. Thus, the velocities inside and outside the SB are given by $|\vec{V}_{SB} - \vec{V}_{HT}| = \alpha V_A$ and $|\vec{V}_{SW} - \vec{V}_{HT}| = \alpha V_A$ (Figure 2b) instead of $V_A$ for the Alfvenic perturbation (Figure 2a). These scalings can be seen in Figure 3b, where $|\vec{V}_{SW} - \vec{V}_{HT}|$ is plotted versus $\alpha V_A$ (the filled circles) and versus $V_A$ (the open circles). $|\vec{V}_{SW} - \vec{V}_{HT}|$ shows better alignment with $\alpha V_A$ than with $V_A$ .

Figure 3c shows the relationship between $|\Delta \vec{V}|$ and the values of $\Delta V_A$ (the red circles) and $V_A$ (the blue circles): the velocity perturbation inside SBs shows a better alignment with $\Delta V_A$ when $\Delta V_A < V_A$ so $\alpha$ can reach 1, and when $\Delta V_A > V_A$ we see that $|\Delta \vec{V}|$ is limited by $V_A$ so $\alpha < 1$. Thus, the value of $|\Delta \vec{V}|$ is limited by the smaller value of $V_A$ and $\Delta V_A$.

Figure 4a presents the dependence of the SBs' Alfvénicity parameter $\alpha$ on the magnetic field deflection angle inside the SBs, $\theta$. The dashed curve presents the upper limit on $\alpha$ based on $\alpha = |B_{SW}|/|\vec{B}_{SB} - \vec{B}_{SW}|$, i.e. $\alpha = 0.5/\sin(\theta/2)$. 17 SBs from 309 detected during Encounter 1 are above this limit, but only 7 of them exceed the limit significantly. The distribution of all perturbations ($\sim 2 \cdot 10^6$ points, corresponding to ~1 measurement per second) in Figure 4b shows that SBs are a natural part of perturbations and that all velocity perturbations are constrained to be smaller than the local Alfvén speed. The distribution of normalized cross helicity $\sigma_c$ inside SBs (Figure 4c) and for all perturbations recorded during Encounter 1 (Figure 4d; where SB points are highlighted by the blue contour) indicates that SBs are closer to field-aligned perturbations (meaning that the angles between $\Delta \vec{V}$ and $\Delta \vec{B}$ are below 20 degrees, and SBs are better tied to the dashed black $\sigma_{c\alpha}$ curve) than general perturbations in the solar wind. The mean value of $\sigma_c$ inside SBs is $0.66 \pm 0.03$, which is lower than the mean value of $\sigma_c$ in the solar wind, $0.69 \pm 0.01$. These estimates correspond well to the time scale decomposed values reported by Bourouaine et al. (2020): $0.6 - 0.7$ inside SBs and $0.75 - 0.85$ for frequencies below $0.05$ Hz. Our estimates are mixed with higher frequency perturbations (frequencies above $0.05$ Hz), where $\sigma_c$ was shown to drop down to $0.2 - 0.4$ inside SBs and slightly lower for solar wind (Bourouaine et al. 2020). We used here the $\sigma_c$ without its scale (or frequency) decomposition to indicate the level of alignment of $\Delta \vec{V}$ and $\Delta \vec{B}$. The distribution of normalized residual density ($\sigma_r = (|\Delta \vec{V}_A|^2 - |\Delta \vec{V}|^2)/(|\Delta \vec{V}_A|^2 + |\Delta \vec{V}|^2$) values is shown in Figure 4e (inside SBs) and Figure 4f (all perturbations from Encounter 1 with SBs marked by the blue contour). SBs-related values are distributed predominantly from -1 to 0 with the mean value of $-0.58 \pm 0.07$ (the mean value of $\sigma_c$ is $-0.42$ based on the data in Figure 4e, which is in a better agreement with the results reported by Bourouaine et al. (2020): $\sigma_c$ varies from $-0.45$ to $-0.3$ inside SBs), the solar wind related intervals demonstrate values distributed around 0 with the mean value of $-0.18 \pm 0.04$.

The observed distribution of $\langle \alpha \rangle_{SB}$ is peaked below one with the maximum around $\alpha = 0.5$-$0.8$. The distribution of time (the number of ~1 s intervals) inside SBs and in the pristine solar wind in panel (f) shows the maximum inside SBs lies between $\alpha = 0.3$ and $\alpha = 0.6$. This difference suggests that the observed duration of SBs with higher $\theta$ is longer than for lower $\theta$, and can be presumably related to the orientation of SBs.

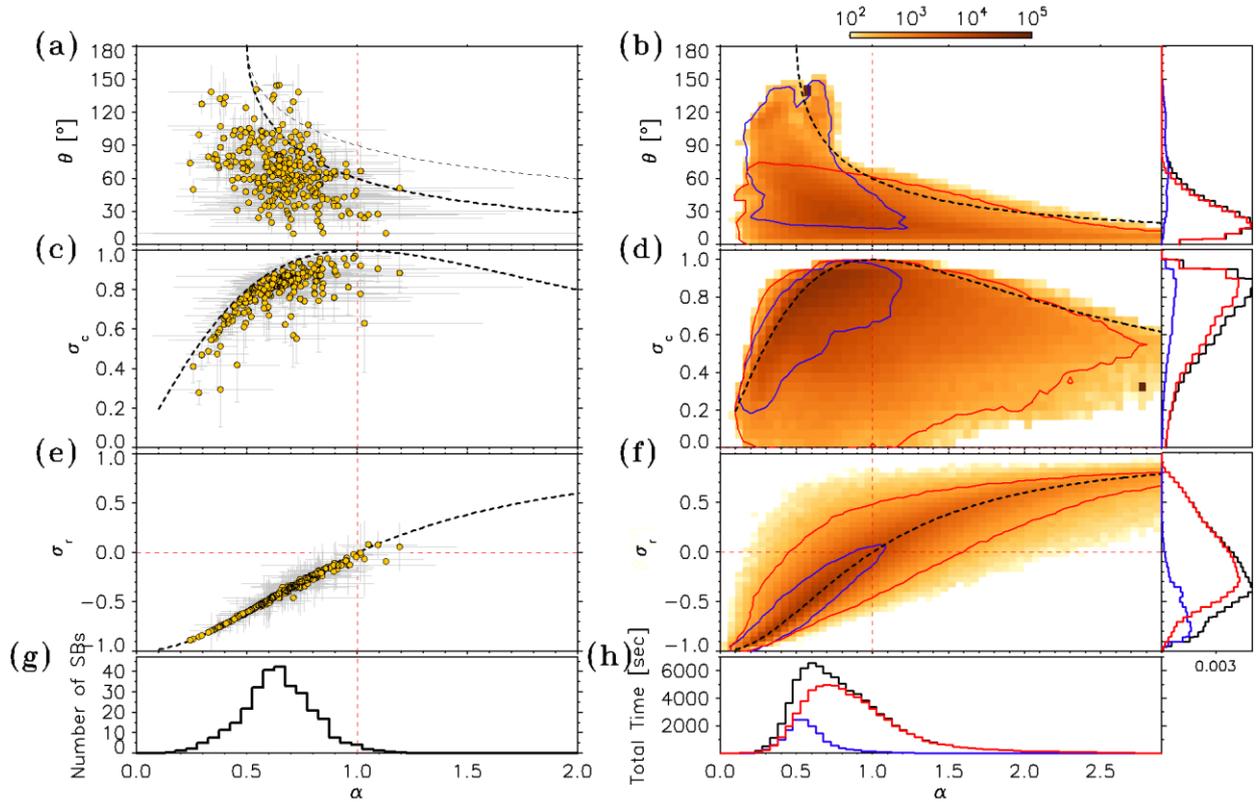

**Figure 4.** The distributions of SBs' averaged parameters (a) - in the $\alpha - \theta$ domain and (b) - the distribution of perturbations of solar wind parameters from Encounter 1 (~$2 \cdot 10^6$ points, corresponding to ~1 measurement per second) in the $\alpha - \theta$ domain (switchback intervals are contoured by blue and non-switchback solar wind intervals are contoured by red; the distribution on $\theta$ is in the right sub-panel with the same color coding). (c) - the distribution of SBs and (d) the distribution of perturbations of solar wind parameters from Encounter 1 in the $\alpha - \sigma_c$ domain (the distribution on $\sigma_c$ is in the right sub-panel). The black dashed curve presents the case of the perfect alignment of $\Delta\vec{V}$ and $\Delta\vec{B}$ when $\sigma_c$ is a function of $\alpha$: $\sigma_{c\alpha} = 2\alpha/(1 + \alpha^2)$. (e) - the distribution of SBs and (f) - the distribution of perturbations of solar wind parameters from Encounter 1 in the $\alpha - \sigma_r$ domain (the distribution on $\sigma_r$ is in the right sub-panel). The black dashed curve presents the case of the perfect alignment of $\Delta\vec{V}$ and $\Delta\vec{B}$ when $\sigma_r$ is a function of $\alpha$: $\sigma_{c\alpha} = (\alpha^2 - 1)/(1 + \alpha^2)$. (g) - the distribution of $\langle\alpha\rangle_{SB}$ values inside SBs (one value from each SB); and (h) - the distribution of $\alpha$ values from solar wind perturbations recorded from Encounter 1 (black), the distribution inside SBs (blue), the distribution in the pristine solar wind (red).

The lower values of $\sigma_r$ inside SBs can appear as the consequence of enhanced Alfvénic turbulence levels (Gogoberidze et al. 2012), which are observed inside SBs at heliocentric distances below ~40 RS (e.g. Dudok de Wit et al.2020; Mozer et al., 2020). Gogoberidze et al. (2012) demonstrated that for stationary critically balanced magnetohydrodynamic turbulence, negative residual energy will always be generated by nonlinear interacting Alfven waves providing negative residual energy (and corresponding lower Alfvénicity) in enhanced solar wind turbulence. The connection of $\alpha$ with $\theta$ indicates that the general limitations on the plasma bulk velocity inside SBs are likely a consequence of surface instabilities such as the Kelvin-Helmholtz instability at the SB boundary, features of which have been observed previously (Mozer et al. 2020; Larosa et al. 2022). Then even for an originally Alfvénic perturbation, $|\Delta\vec{V}|$ is relaxed to the value below the local $V_A$ to sustain locally the perturbation surface stability, so decreasing SB's Alfvénicity during propagation.

## 3. CONCLUSIONS

Based on Parker Solar Probe observations, we examined the perturbations of magnetic field and plasma bulk velocity inside switchbacks (SBs) observed during Encounter 1 and quantified the deviations from an ideal Alfvénic relationship (estimated from the magnetic field perturbation inside the SB). We have found that:

1. The cross-helicity values inside SBs confirm the better alignment (the angle between the velocity and magnetic field is below 20 degrees) of the magnetic field and plasma bulk velocity perturbations inside SBs compared with that in the pristine solar wind;

2. SBs and the local solar wind have well-defined, common de Hoffman-Teller frames. In this frame the velocity difference between the axial plasma flow inside a SB (along the magnetic field) and the ambient solar wind velocity were expected to equal the corresponding Alfvén speed difference $\Delta\vec{V}_A = \Delta\vec{B}/\sqrt{4\pi n_i m_i}$.. However, the data reveal a systematic deficit of perturbation velocity magnitudes compared to the corresponding Alfvénic perturbations estimated based on the magnetic field perturbation inside the SB by the minimal of $\Delta\vec{V}_A$ or $V_A$.

3. The parameter $\alpha = |\Delta\vec{V}|/|\Delta\vec{V}_A|$ is used to quantify the deviation from the ideal Alfvénic condition ($\alpha = 1$). The data reveal a clear dependence of $\alpha$ on the perturbation deflection angle $\theta$ with an upper limit on $\alpha$ controlled by the SB deflection angle $\theta$ ($\alpha < 0.5/\sin(\theta/2)$ or $\alpha < |B_{SW}|/|\vec{B}_{SB} - \vec{B}_{SW}|$). This limit follows from the requirement that the SB boundary velocity shear $|\Delta\vec{V}|$ not exceed the local Alfvén speed ($V_A$). Thus, the perturbations of the proton bulk velocity $|\Delta\vec{V}|$ in SBs with $\theta > \pi/3$ cannot reach $\alpha = 1$ (the Alfvénicity condition for the perturbation), and the highest possible $\alpha$ for a SB that contains full reversal of the magnetic field ($\theta = \pi$) is 0.5.

An important question is whether the observational evidence that the Alfvénicity parameter is less than unity for SBs with large rotation angles has implications for models of switchback formation. Specifically, do the SB models based on the steepening of Alfvén waves during the expansion of the radial magnetic field with distance from the sun (e.g. Squire et al 2020, 2022) also reveal reduced Alfvénicity in SBs with large rotations? Simulations can hopefully shed light on these new constraints based on the PSP observations.

## AKNOWLEDGEMENT

The work of was supported by NASA grants 80NSSC22K0433, 80NNSC19K0848. The work of OVA and KEC was partially supported by NSF grant number 1914670 and NASA grants contracts 80NSSC22K0522, 80NSSC20K0697 and NASA's Living with a Star (LWS) program (contract 80NSSC20K0218). We thank the NASA Parker Solar Probe Mission, SWEAP team led by Justin Kasper, and FIELDS team led by Stuart Bale for the use of the data. The FIELDS experiment on the Parker Solar Probe spacecraft was designed and developed under NASA contract NNN06AA01C.